\documentclass[twocolumn]{aastex631}
\usepackage[graphicx]{realboxes}
\usepackage{txfonts}
\usepackage{newtxtext}
\usepackage{mathrsfs}
\usepackage{fourier}
\usepackage[T1]{fontenc}
\usepackage{xcolor}
\usepackage{color}
\usepackage{soul}
\usepackage{float}
\usepackage{hyperref}
\usepackage{rotating}
\usepackage{comment}
\usepackage[toc,page]{appendix}

\begin{document}

\title{COSMIC-L: A Photometric Catalog of Observed Stars in the Large MagellanIc Cloud}

\correspondingauthor{Antonio Franco}
\email{antonio.franco@le.infn.it}

\author[0000-0002-4761-366X]{Antonio Franco}
\affiliation{INFN, Sezione di Lecce, Via per Arnesano, CP-193, I-73100, Lecce, Italy}
\affiliation{INAF, Sezione di Lecce, Via per Arnesano, CP-193, I-73100, Lecce, Italy}
\affiliation{Department of Mathematics and Physics {\it ``E. De Giorgi''}, University of Salento, \\ Via per Arnesano, CP-I93, I-73100, Lecce, Italy}

\author[0000-0002-7926-3481]{Achille A. Nucita}
\affiliation{INFN, Sezione di Lecce, Via per Arnesano, CP-193, I-73100, Lecce, Italy}
\affiliation{INAF, Sezione di Lecce, Via per Arnesano, CP-193, I-73100, Lecce, Italy}
\affiliation{Department of Mathematics and Physics {\it ``E. De Giorgi''}, University of Salento, \\ Via per Arnesano, CP-I93, I-73100, Lecce, Italy}

\author[0000-0001-6460-7563]{Francesco De~Paolis}
\affiliation{INFN, Sezione di Lecce, Via per Arnesano, CP-193, I-73100, Lecce, Italy}
\affiliation{INAF, Sezione di Lecce, Via per Arnesano, CP-193, I-73100, Lecce, Italy}
\affiliation{Department of Mathematics and Physics {\it ``E. De Giorgi''}, University of Salento, \\ Via per Arnesano, CP-I93, I-73100, Lecce, Italy}

\author[0000-0002-8757-9371]{Francesco Strafella}
\affiliation{INFN, Sezione di Lecce, Via per Arnesano, CP-193, I-73100, Lecce, Italy}
\affiliation{INAF, Sezione di Lecce, Via per Arnesano, CP-193, I-73100, Lecce, Italy}
\affiliation{Department of Mathematics and Physics {\it ``E. De Giorgi''}, University of Salento, \\ Via per Arnesano, CP-I93, I-73100, Lecce, Italy}

\begin{abstract}

The Magellanic Clouds are two nearby dwarf irregular galaxies whose study can help us in understanding galaxy and stellar evolution. In particular, the Large Magellanic Cloud, the larger one, contains approximately 30 billion stars at various evolutionary stages. In this work, we present an SDSS $gri$-bands photometric analysis based on multiple images acquired with DECam, the Dark Energy Camera, installed on the Blanco telescope at Cerro Tololo Inter-American Observatory (Chile). We performed a full image analysis and photometric calibration, resulting in a photometric catalog named COSMIC-L, consisting of 57,997,665 stars, of which 18,676,294 contain estimates for all three $gri$ magnitudes, resulting in a completeness magnitude of $\simeq 21$ and a limiting magnitude of $\simeq 22$ in all three bands.

\end{abstract}

\keywords{catalogs --- techniques: photometric --- stars: general --- galaxies: Magellanic Clouds }


\section{Introduction}

Stellar photometry is one of the most fundamental techniques in astronomy for characterizing stellar evolutionary states. his can be achieved by analyzing the source’s light filtered at different wavelengths, providing crucial information about their physical properties, such as temperature, composition, age, and distance. The Large Magellanic Cloud (LMC hereafter), a nearby dwarf satellite galaxy of the Milky Way, is an ideal target for photometric studies. Due to its proximity, it offers a unique opportunity to observe stellar populations in great detail. Photometry, i.e. the study of light coming from a certain source, plays a key role in understanding the formation and evolution of stars within the LMC, as well as in refining models of stellar evolution. Many surveys targeting the LMC and its smaller twin galaxy, the Small Magellanic Cloud (SMC), have been conducted to study the local stellar populations. Among these, OGLE, EROS, and MACHO are representative examples of such surveys. Moreover, photometry in the LMC is essential for calibrating the extragalactic distance scale. Since the LMC hosts well-known standard candles, such as RR Lyrae stars and Cepheids (see, e.g., \citealt{franco2023, soszynki2015a, soszynki2015b, soszynki2016, soszynki2017, soszynki2018, udalski2015}), precise photometric measurements help improve our estimates of cosmic distances. This, in turn, contributes to a more accurate determination of the Hubble constant, which describes the expansion rate of the universe. Furthermore, the high density of sources in the LMC region, as well as in the SMC, permits the detection of transients such as microlensing events (see, e.g., \citealt{calchinovati2013, franco2024, mroz2024a, mroz2024b}) that can probe the presence of dark matter objects in the Milky Way's galactic halo.

In this work, we present COSMIC-L, a {\it Photometric Catalog of Observed Stars in the Large MagellanIc Cloud}, containing more than 57 million sources in the LMC field, obtained by analyzing 23 pointings with DECam, the Dark Energy Camera \citep{flaugher2015}, adopting the analysis strategy described in \cite{franco2025} for the COSMIC-S catalog, where it was applied to stars in the SMC.
In Section \ref{Sec::DataReduction} we provide an overview of DECam, describe the dataset used, and explain the data reduction process.
In Section \ref{Sec::catalog} we introduce the COSMIC-L catalog, presenting the obtained results. Final conclusions are presented in Section \ref{Sec::conclusion}.

\begin{figure*}[ht!]
    \centering
    \includegraphics[width=0.49\textwidth]{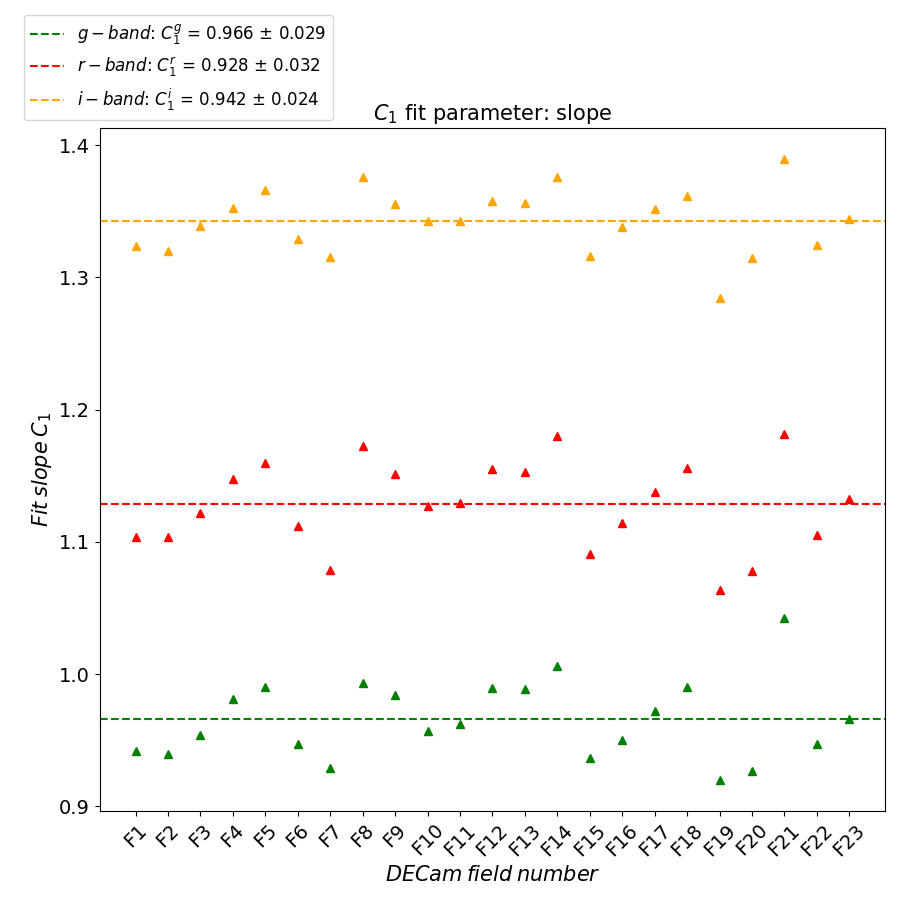}
    \includegraphics[width=0.49\textwidth]{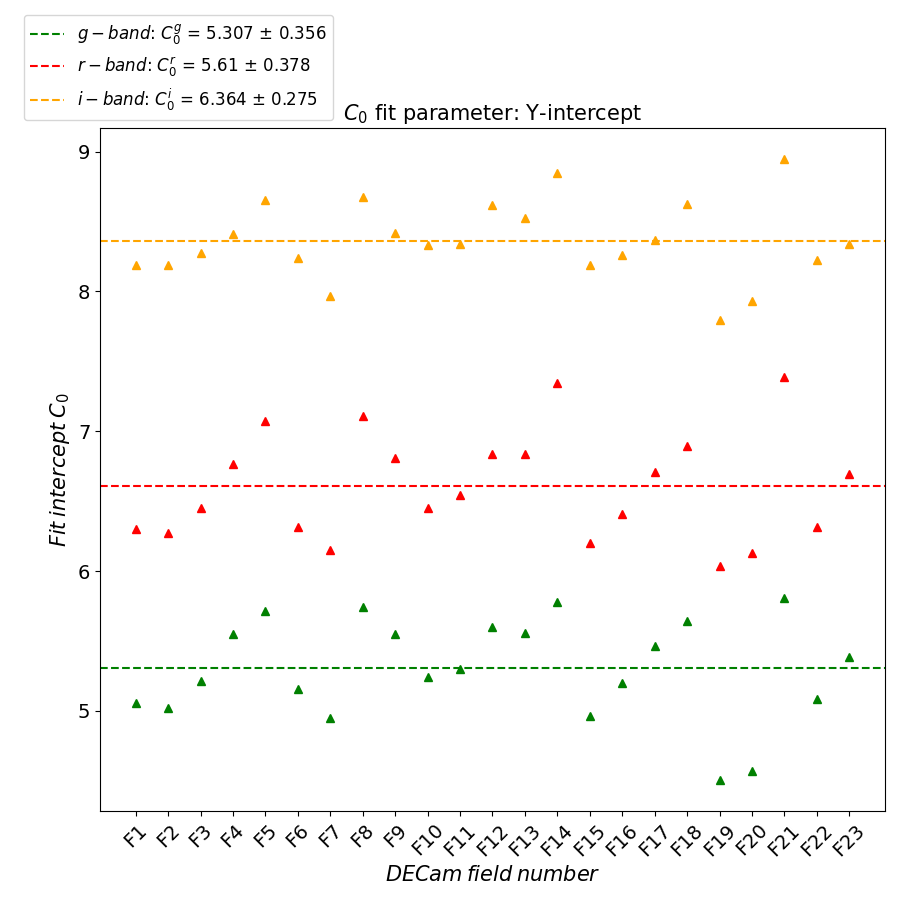}
    \caption{Slope (left panel, named $C_1$) and Y-intercept (right panel, named $C$) distribution obtained after the photometric calibration fit for all three photometric bands, specifically $g$-band in green, $r$-band in red and $i$-band in orange. The two values have been provided by the photometric calibration fit, i.e. $m_{ATLAS}=C_1\>m_{DECam}+C_0$. On the x-axis the investigated DECam fields are reported labeled as {\it F*}, where the asterisk stands for the number of the field, from 1 to 23. The y-axis shows the slope and Y-intercept values to which the calibration fit converged. The dashed lines indicate the mean values for each band. In order to make the figure readable, a vertical shift of 0.2 and 1.0, for the slope and Y-intercept distributions respectively, has been applied for the $r$-band and $i$-band distribution, taking the $g$-band values as reference. Each panel includes, in the upper left box, the mean values and corresponding uncertainties, calculated as the standard deviation, for each filter.}%
    \label{Fig::calib_fit}%
\end{figure*}

\section{\label{Sec::DataReduction} Observations and data reduction}

Observations were conducted using DECam on the 4m V. Blanco Telescope at CTIO (Chile) with the $g$, $r$, and $i$ SDSS filters \citep{lenz1998}. DECam consists of 62 scientific CCDs, each with a resolution of 0.27''/pixel, covering approximately 3.0 deg$^2$ of the sky \citep{honscheid2008}. The data used in this study were obtained from the DECam program 2018A-0273 (February 2018 – January 2020) and are publicly accessible via the NOIRLab Archive. Observations covered the LMC region through 23 different DECam fields in the $gri$ bands, ensuring nearly complete coverage of the LMC, with only minor gaps that were subsequently addressed.Exposure times were 200, 100, and 200 seconds per band, respectively.

Images were reduced through bias, dark, and flat-field corrections, and then processed using a custom Python pipeline. External software tools, such as SExtractor \citep{sextractor} and PSFEx \citep{psfex}, were used for source detection and PSF extraction, as PSF photometry is essential in crowded stellar fields, unlike aperture photometry.
 
We first used SExtractor to identify isolated and bright sources in the DECam images. We estimated the PSF of the stars using PSFEx, then reprocessed the source catalog to obtain more accurate photometry. We then extracted the sources' instrumental magnitudes, which were calibrated using the ATLAS All-Sky Stellar Reference Catalog \citep{tonry2018} as the photometric reference. We applied a robust linear regression to correct the instrumental magnitudes, providing a photometric calibration for each DECam field and photometric band.

The resulting fit parameters are summarized in Figure~\ref{Fig::calib_fit}, showing the slopes (left panel) and Y-intercepts (right panel) for each DECam field, labeled from F1 to F23, as obtained from the photometric calibration fit in which the DECam magnitudes and the reference ATLAS magnitudes have been correlated by using the standard relation $m_{ATLAS} = C_1\>m_{DECam} + C_0$, where the ATLAS and DECam magnitudes are considered for each filter used, while the $C_1$ and $C_0$ parameters represent the fit slope and intercept, respectively. The three colors refer to the considered photometric band, i.e. $g$-band in green, $r$-band in red, and $i$-band in orange. 
The dashed lines indicate the mean values for each band. A vertical shift of 0.2 and 1.0 between the $g$-band and $r$-band sample, and the $r$-band and $i$-band sample, respectively, has been applied in order to make the figure easier to read. Note that for both the slopes and the Y-intercepts, the values obtained for the three filters show similar behavior. This is because the images were taken over the same 23 fields using all three $gri$ filters. As a result, the calibration was influenced by the same boundary conditions, such as stellar crowding. However, this does not affect the quality of the photometric calibration, which was performed independently for each field using its specific calibration parameters, as shown in Figure~\ref{Fig::calib_fit}. Finally, we applied the calibrated PSF photometry to all detected sources using the extracted PSF models. Sources with poorly fitted PSFs or other anomalies were removed to produce the final version of the catalog. The detailed analysis procedure is discussed thoroughly in \cite{franco2025}.

\begin{table*}
    \centering
    \begin{tabular}{llll} 
        \hline \hline
        \textbf{Column} & \textbf{Column} & \textbf{Units} & \textbf{Description}  \\
        \textbf{Number} & \textbf{Name} & &  \\
        \hline
        1 & COSMIC-L\_ID &  & Object Identification Number \\
        2 & RAJ2000 & deg & Right Ascension associated to the selected source \\
        3 & DEJ2000 & deg & Declination associated to the selected source \\
        4 & e\_RAJ2000 & deg & Uncertainty associated to the Right Ascension \\
        5 & e\_DEJ2000 & deg & Uncertainty associated to the Declination \\
        6 & gmag & mag & SDSS g-band magnitude \\
        7 & e\_gmag & mag & Uncertainty associated to the SDSS g-band magnitude \\
        8 & rmag & mag & SDSS r-band magnitude \\
        9 & e\_rmag & mag & Uncertainty associated to the SDSS r-band magnitude \\
        10 & imag & mag & SDSS i-band magnitude \\
        11 & e\_imag & mag & Uncertainty associated to the SDSS i-band magnitude \\
        \hline \hline
    \end{tabular}
    \caption{Overview of the columns in the COSMIC-L photometric catalog. Column 1 reports the column number; column 2 shows the corresponding catalog field name; column 3 provides the physical units; column 4 describes each parameter.}
    \label{Table::cat_info}
\end{table*}

\begin{figure}[!htp]
    \centering
    \includegraphics[width=\columnwidth]{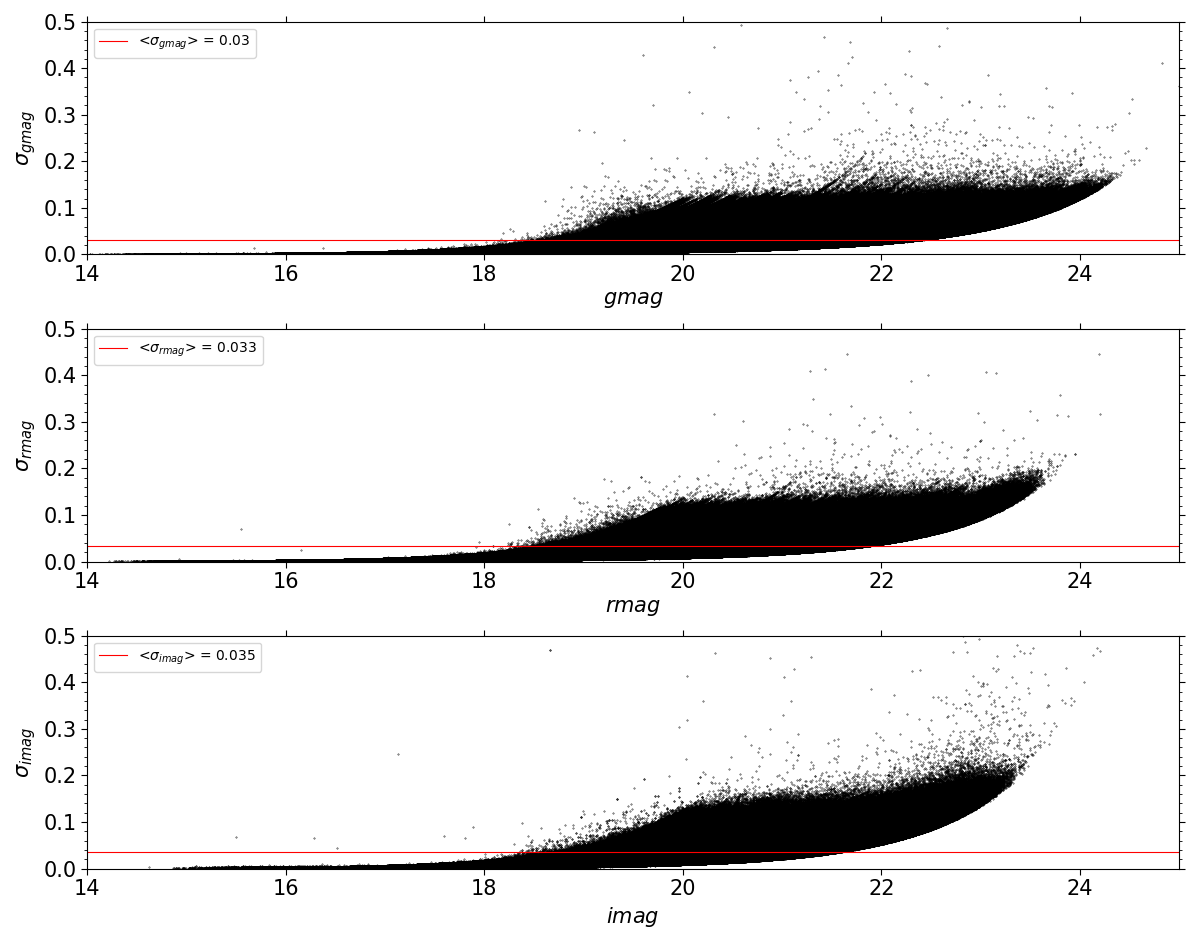}%
    \caption{Photometric uncertainties as a function of magnitude in the $g$, $r$, and $i$ bands (top to bottom) for all sources in the COSMIC-L catalog. The red horizontal line indicates the mean photometric error in each band. Photometry is based on PSF fitting.}
    \label{Fig::mag_vs_err}%
\end{figure}


\section{\label{Sec::catalog} The COSMIC-L photometric catalog}

The final catalog, named COSMIC-L, includes 57,997,665 sources, of which 18,676,294 exhibit reliable photometry in all three $gri$ bands. For each entry, we provide an identification number (Column 1), celestial coordinates and their uncertainties in the J2000 reference frame (Columns 2–5), and calibrated magnitudes in the $g$, $r$, and $i$ bands with the corresponding photometric errors (Columns 6–11). A complete description of the catalog structure is provided in Table~\ref{Table::cat_info}.

\begin{figure}[ht!]
    \centering
    \includegraphics[width=\linewidth]{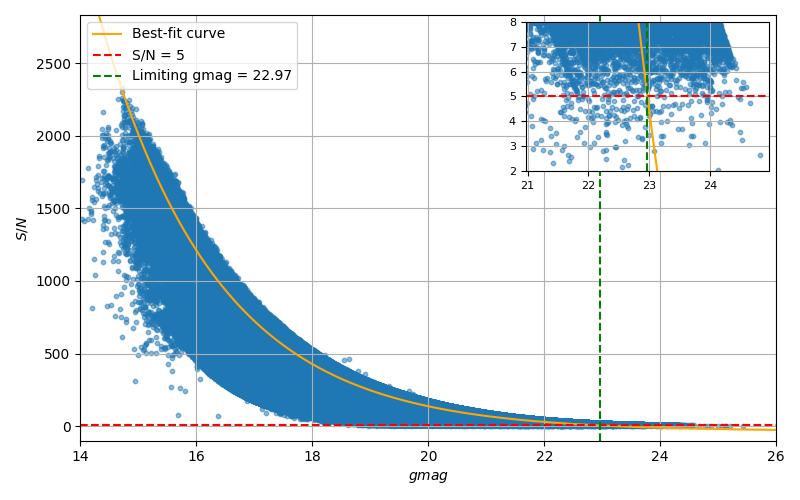}
    \includegraphics[width=\linewidth]{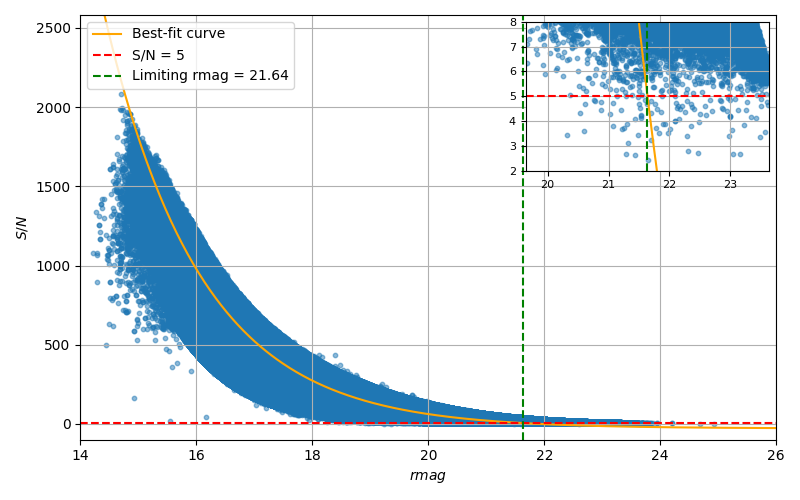}
    \includegraphics[width=\linewidth]{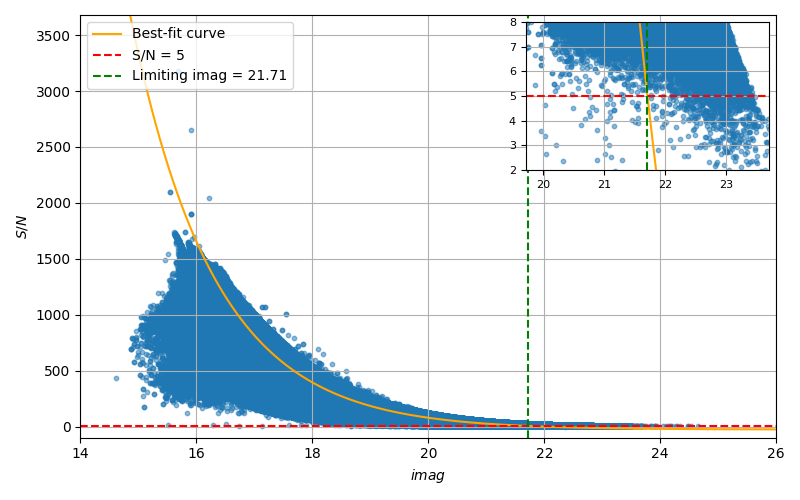}
    \caption{Signal-to-noise ratio (SNR) as a function of magnitude in the $g$ (top), $r$ (middle), and $i$ (bottom) bands. The data are fitted with a decaying exponential function, shown as a yellow line. The adopted SNR threshold ($S/N = 5$) is marked by the red horizontal dashed line, while the corresponding limiting magnitude is indicated by the green vertical dashed line. Each panel includes an upper-right sub-panel that zooms into the region around the limiting magnitude.}
    \label{Fig::limiting}
\end{figure}

\begin{figure*}[ht!]
    \centering
    \includegraphics[width=0.49\textwidth]{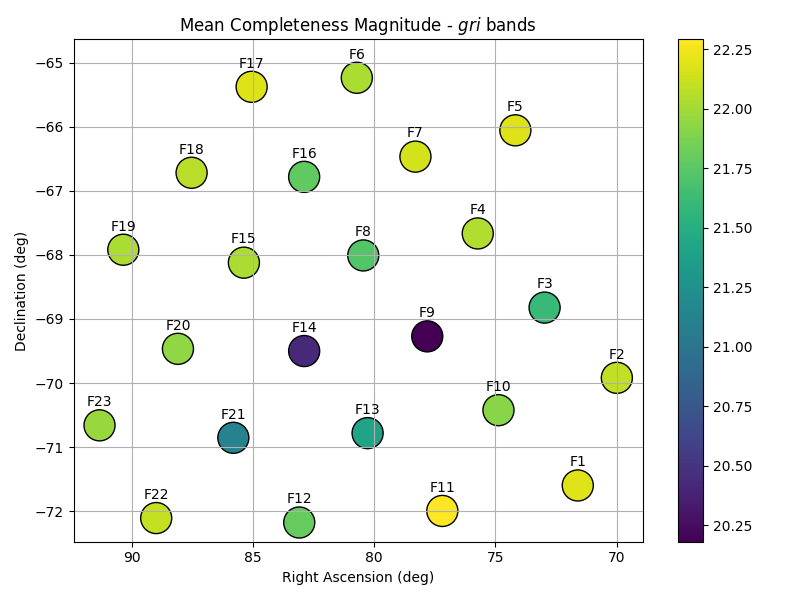}
    \includegraphics[width=0.49\textwidth]{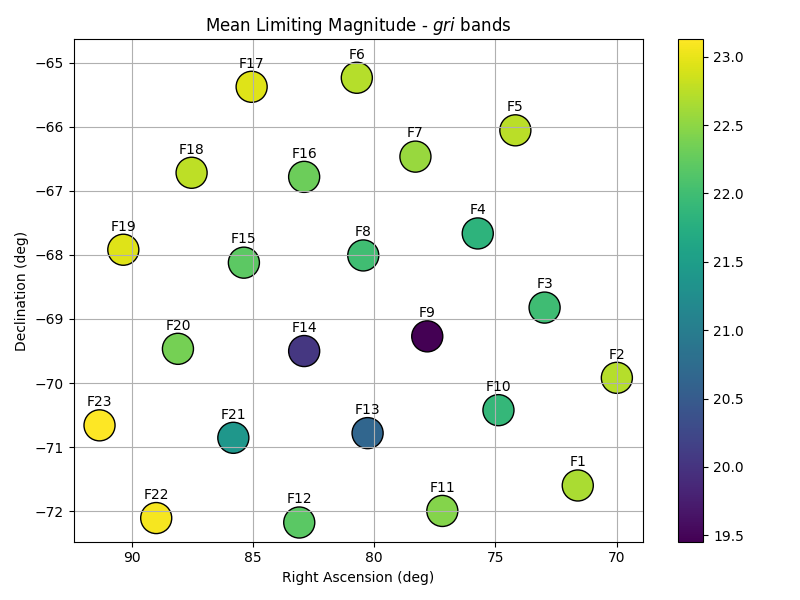}
    \caption{Spatial distribution of the completeness (left panel) and limiting (right panel) magnitudes across the LMC region, based on the COSMIC-L catalog. The color bars indicate the magnitude values for each of the 23 DECam fields, labeled according to the field numbering convention (i.e., F1 to F23). See the text for further discussion.}%
    \label{Fig::mag_spatial_distr}%
\end{figure*}

\begin{figure}[!htp]
    \centering
    \includegraphics[width=\columnwidth]{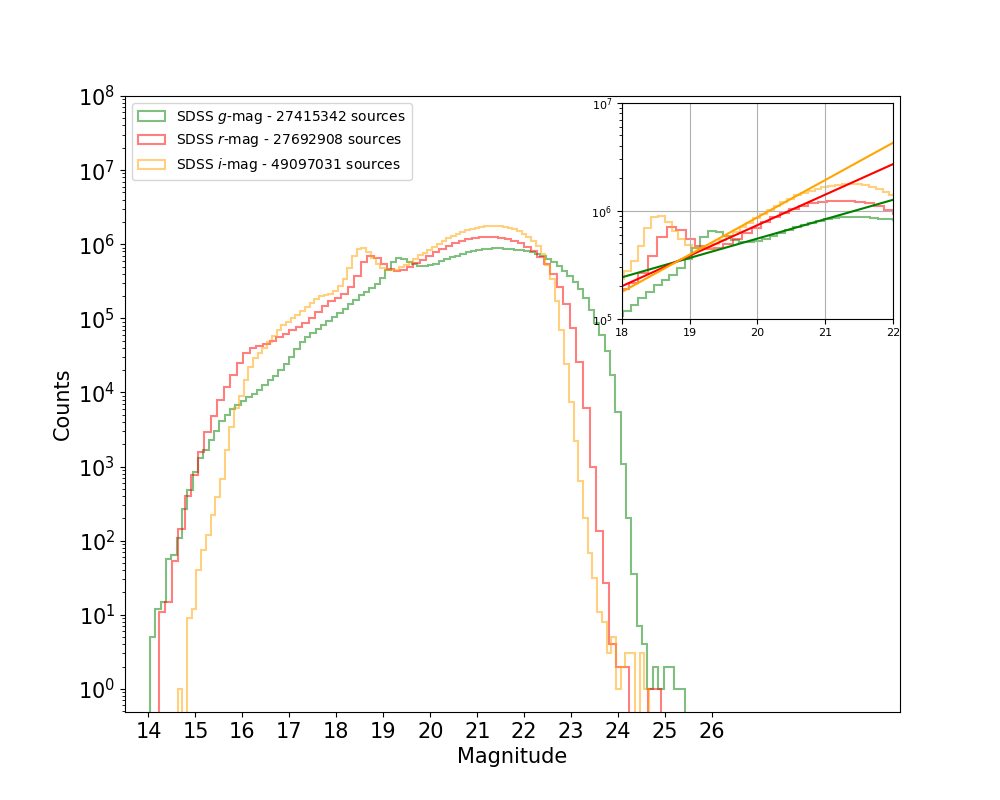}%
    \caption{
    Magnitude distributions of the detected sources in the three photometric bands used in this study: $g$ (green), $r$ (red), and $i$ (yellow). The legend in the top-left corner indicates the number of detections per band. The upper-right sub-panel shows a zoom-in around the histogram peaks, along with linear fits (colored lines) used to estimate the completeness magnitudes. For clarity, the resulting completeness values are listed in Table~\ref{Table::compl_lim}.}
    \label{Fig::mag_hist}%
\end{figure}

\begin{table}[!htp]
    \centering
    \begin{tabular}{ll} 
        \hline \hline
        \textbf{Data product} & \textbf{Number of sources} \\
        \hline
        Total sources & 57~997~665 \\
        $g$-band magnitude &  27~415~342 \\
        $r$-band magnitude &  27~692~908 \\
        $i$-band magnitude &  49~097~031 \\
        $gri$ mag & 18~676~294 \\
        \hline \hline
    \end{tabular}
    \caption{Total number of sources in the COSMIC-L catalog, including both single-band detections and sources with photometry in all three $gri$ bands.}
    \label{Table::number_of_products}
\end{table}

\begin{figure*}[ht!]
    \centering
    \includegraphics[width=0.49\textwidth]{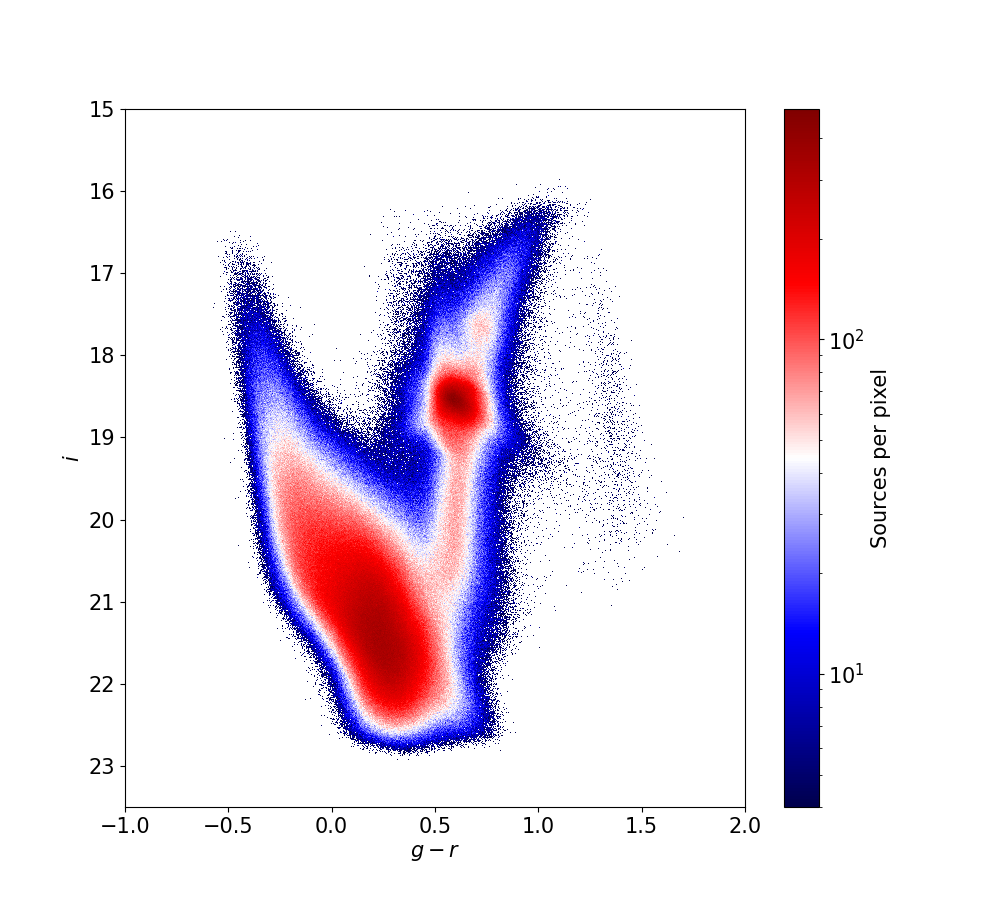}
    \includegraphics[width=0.49\textwidth]{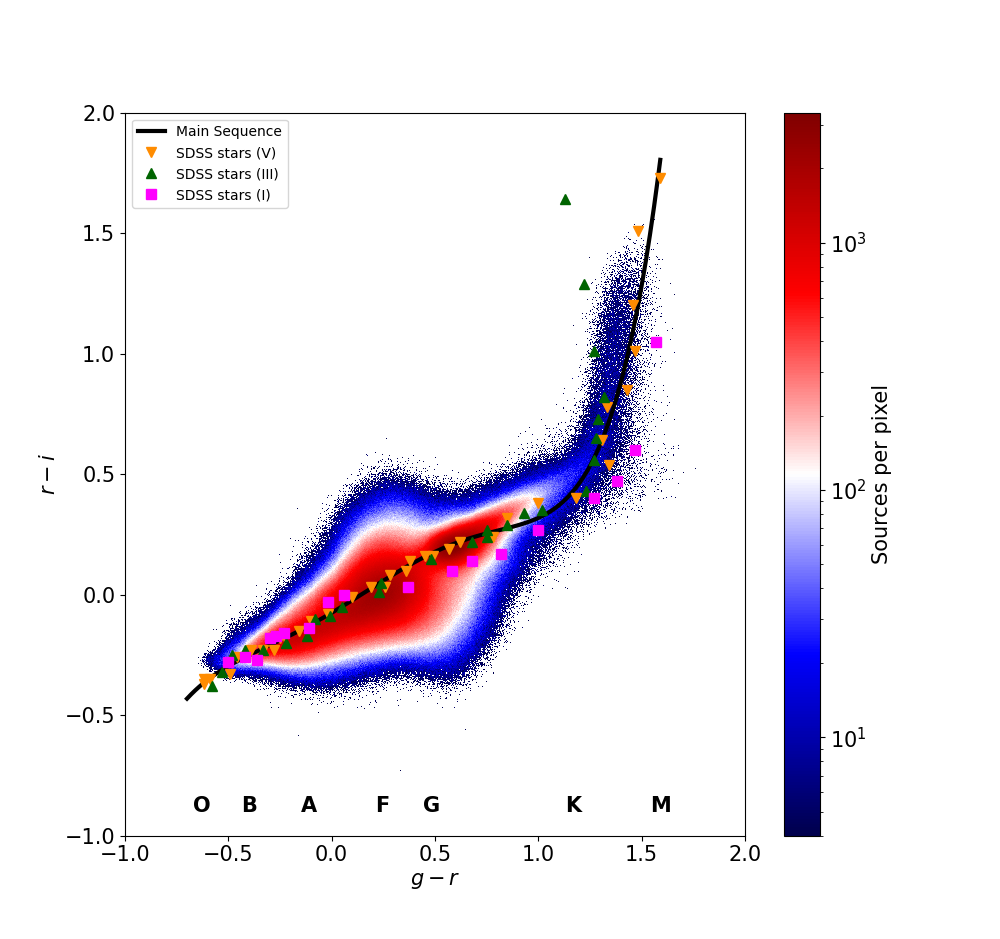}
    \caption{
    Left panel: color–magnitude diagram ($i$ vs. $g - r$) for LMC sources, highlighting the main sequence (left) and the giant branch (right). A dense population with $g - r > 1.2$ corresponds to foreground K-type and M-type stars in the Milky Way. Right panel: color–color diagram ($r - i$ vs. $g - r$) for the same sources. Spectral types (O, B, A, F, G, K, M) are labeled along the bottom, while luminosity classes (V, III, I) are shown in orange, green, and fuchsia, respectively. The main sequence is marked by a black line. A total of 18,676,294 sources are included. For details, see Table~\ref{Table::number_of_products}. Both diagrams use 1000-bin 2D histograms, discarding bins with fewer than 3 sources to reduce noise. The color bars indicate source density per pixel.}
    \label{Fig::hr_diag}
\end{figure*}

\begin{figure}[ht!]
    \centering
    \includegraphics[width=\columnwidth]{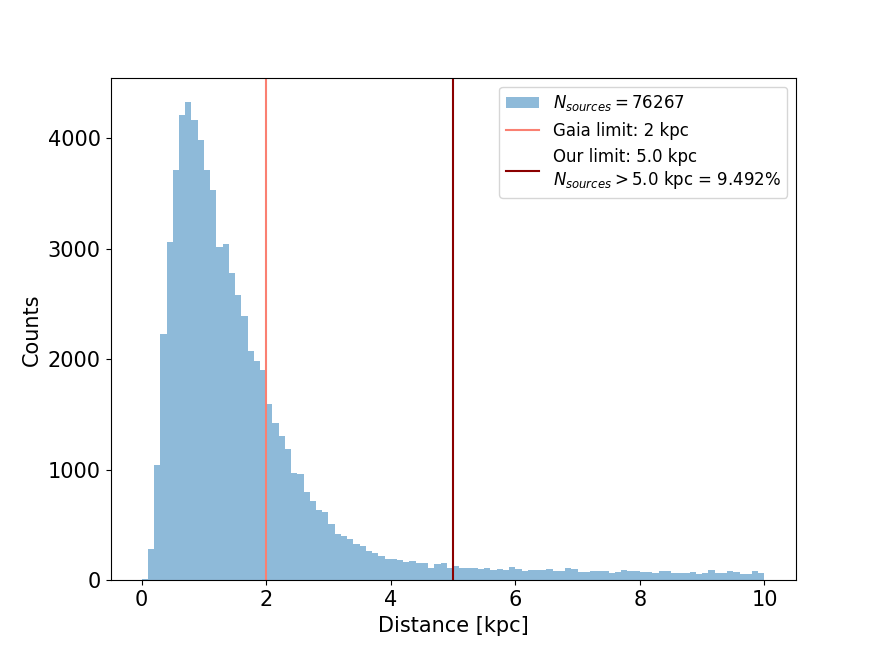}%
    \caption{
    Distance distribution of the 76,267 stars with $g - r > 1.2$, identified in Figure~\ref{Fig::hr_diag}. Distances are derived from Gaia EDR3, with corrections applied using \cite{bailerjones2021}. The Gaia nominal detection limit at $\sim$2kpc is marked by the pink vertical line \citep{sanchez2010}, while the extended detection limit adopted in this work is indicated by the red vertical line. Approximately 91\% of the sources lie within 5kpc, confirming their classification as foreground stars in the thin and thick disks of the Milky Way. The top-right inset contains the corresponding legend.}
    \label{Fig::dist_hist}
\end{figure}

Positional uncertainties, computed by SExtractor, are directly included in the COSMIC-L catalog, with both coordinates and errors expressed in degrees. Photometric magnitudes result from the calibration of DECam instrumental measurements—extracted using SExtractor and PSFEx—against the ATLAS All-Sky Stellar Reference Catalog. The final magnitude uncertainties are calculated as the quadrature sum of the relevant error terms, following the approach described in \citet{franco2025}.

Figure~\ref{Fig::mag_vs_err} shows the distribution of photometric errors as a function of magnitude for the $g$, $r$, and $i$ bands. The red horizontal lines indicate the average uncertainties in each band: $<\sigma_{gmag}> = 0.03$, $<\sigma_{rmag}> = 0.033$, and $<\sigma_{imag}> = 0.035$. The corresponding magnitude distributions for the three filters -- $g$ (green), $r$ (red), and $i$ (yellow) -- are presented in Figure~\ref{Fig::mag_hist}.

Based on the data shown in Figure~\ref{Fig::mag_vs_err} and Figure~\ref{Fig::mag_hist}, we estimate the completeness and limiting magnitudes by adopting a widely used method in the absence of injection–recovery simulations. To estimate the completeness magnitude, we performed a linear fit to the magnitude histogram just before its peak. For each band, we identified the magnitude at which the observed counts deviate by 5\% from the extrapolated linear fit. This procedure is illustrated in the upper-right sub-panel of Figure~\ref{Fig::mag_hist}, where the colored lines, matching the histogram colors, represent the linear fits. The resulting completeness magnitudes are 21.38, 20.82, and 20.79 for the $g$, $r$, and $i$ bands, respectively (see first column of Table~\ref{Table::compl_lim}).
To estimate the limiting magnitude, we derived the signal-to-noise ratio (SNR) using the standard relation:
\begin{equation}
    S/N = \frac{2.5}{\ln{10} \cdot \sigma_m}
\end{equation}
where $\sigma_m$ is the photometric uncertainty. We then extrapolated the magnitude corresponding to a detection threshold of $S/N = 5$, a commonly adopted criterion for reliable source detection. After computing the SNR values for all sources, we fitted an exponential decay function to the data, as shown in the three panels of Figure~\ref{Fig::limiting}. In each plot, the best-fit curve is shown in yellow, the limiting magnitude is marked by a green vertical dashed line, and the SNR threshold is indicated by a red horizontal dashed line. The derived limiting magnitudes are 22.97, 21.64, and 21.71 for the $g$, $r$, and $i$ bands, respectively (see second column of Table~\ref{Table::compl_lim}).

Although based on indirect methods, the estimation of completeness and limiting magnitudes provides a consistent and reproducible approach to assess the depth and reliability of the catalog. Table~\ref{Table::compl_lim} summarizes these values, showing that the catalog is complete down to $m_c \lesssim 21$ and reaches a detection limit of $m_l \simeq 22$ across all three bands. Nonetheless, numerous sources are detected down to magnitudes as faint as $m \simeq 25$, highlighting the remarkable performance achieved with ground-based observations.

We also analyzed the spatial distribution of the completeness and limiting magnitudes across the 23 DECam fields, as shown in Figure~\ref{Fig::mag_spatial_distr}. The left and right panels display the completeness and limiting magnitudes, respectively, with the color bars indicating the magnitude values across the LMC field. As expected, the central regions of the cloud, specifically fields F9 and F14, show lower completeness and limiting magnitudes due to the higher stellar density, whereas the outer regions reach fainter limits, with magnitudes approaching $\simeq22$. The global values reported in Table~\ref{Table::compl_lim}, computed starting from the whole dataset, can also be interpreted as weighted averages, where the weights are proportional to the number of sources in each field.

Figure~\ref{Fig::hr_diag} shows the color–magnitude diagram ($i$ vs. $g - r$) and the color–color diagram ($r - i$ vs. $g - r$) for sources in the LMC. Both the main sequence and the giant branch are clearly visible. The densest and reddest region, located at $g - r \simeq 0.6$ and $i \simeq 18.5$, corresponds to the LMC Red Clump stars. In addition, a distinct group of sources at $1.2 \lesssim g - r \lesssim 1.6$ includes 76,267 stars, identified as foreground K- and M-type stars belonging to the Milky Way. Their distances, derived from Gaia EDR3 \citep{gaiadr3} and corrected using the method of \cite{bailerjones2021}, are presented in Figure~\ref{Fig::dist_hist}. Approximately 90\% of these stars lie within 5~kpc, confirming their location in the Galactic thin and thick disks.

The color–color diagram (right panel of Figure~\ref{Fig::hr_diag}) further supports this classification. The Red Clump is centered at $g - r \simeq 0.5$ and $r - i \simeq 0.2$, while a noticeable change in slope for $g - r \gtrsim 1.2$ marks the transition to K-type and M-type stars. Spectral types (O, B, A, F, G, K, M) and luminosity classes (V, III, I) from \citet{covey2007} are also plotted, with the main sequence traced in black.

\section{\label{Sec::conclusion} Conclusions}

In this work, we presented COSMIC-L, a deep photometric catalog containing more than 57 million sources in the direction of the Large Magellanic Cloud. The data, obtained with the DECam instrument in the $g$, $r$, and $i$ SDSS filters, were processed with a custom pipeline based on SExtractor and PSFEx, enabling accurate PSF photometry. Among the total sources, nearly 19 million have a good photometry in all three bands. For these, we constructed and analyzed the corresponding color–magnitude and color–color diagrams, highlighting the different stellar populations. In particular, the color–magnitude diagram allowed us to identify a prominent foreground population of K-type and M-type stars in the Milky Way, which appears as a vertical feature at $g - r \simeq 1.5$. The COSMIC-L catalog is virtually complete down to a magnitude of $m_c \simeq 21$, with a detection limit reaching $m_l \simeq 22$. This makes COSMIC-L the deepest ground-based photometric catalog of the LMC stars currently available.

\begin{table}[h!]
    \centering
    \begin{tabular}{lll} 
        \hline \hline
        \textbf{Magnitude} & \textbf{Completeness} & \textbf{Limiting} \\
         \textbf{band} & \textbf{magnitude} & \textbf{magnitude} \\
        \hline
        $g$-mag & 21.38 & 22.97 \\
        $r$-mag & 20.82 & 21.64 \\
        $i$-mag & 20.79 & 21.71 \\
        \hline \hline
    \end{tabular}
    \caption{Completeness and limiting magnitudes for each photometric band analyzed in this study. Completeness magnitudes are estimated from the magnitude histograms in Figure~\ref{Fig::mag_hist} as the points where the observed distributions deviate by more than 5\% from the linear trend. Limiting magnitudes correspond to the magnitude at which a $S/N = 5$ threshold is reached, as derived from the signal-to-noise fits shown in Figure~\ref{Fig::limiting}.}
    \label{Table::compl_lim}
\end{table}

\section*{Acknowledgements}

This paper is based on publicly available observations by DECam (Dark Energy Camera), an instrument mounted on the V. Blanco Telescope, as part of the Cerro Tololo Inter-American Observatory (Chile). DECam images used for this work are publicly available at the \url{https://astroarchive.noirlab.edu/portal/search/} webpage. We thank the INFN projects ICSC - National Centre for HPC, Big Data and Quantum Computing. We also thank for partial support the INFN projects Euclid and TAsP. This work has been partly supported by the MUR grant PRIN 2022383WFT.


\end{document}